\documentclass[authoryear,preprint,review,12pt]{elsarticle}

\usepackage{graphicx}
\usepackage{amssymb}
\usepackage{color}
\usepackage{longtable}
\usepackage[small,bf,singlelinecheck=off]{caption}
\long\def\comment#1{}

\journal{Icarus}

\begin{document}

\begin{frontmatter}



\title{Lightcurves of the Karin family asteroids}


\author[nao]{Fumi Yoshida}\ead{fumi.yoshida@nao.ac.jp}
\author[nao]{Takashi Ito}
\author[bit]{Budi Dermawan}
\author[htd]{Tsuko Nakamura}
\author[nro]{Shigeru Takahashi}
\author[uba]{Mansur A. Ibrahimov}
\author[lpl]{Renu Malhotra}
\author[ncu]{Wing-Huen Ip}
\author[ncu]{Wen-Ping Chen}
\author[fue]{Yu Sawabe}
\author[fue]{Masashige Haji}
\author[fue]{Ryoko Saito}
\author[fue]{Masanori Hirai}
%\author[tmo]{Seidai Miyasaka}
%\author[nao]{Hideo Fukushima}
%\author[nao]{Hideo Sato}
%\author[hkd]{Yusuke Sato}

%\author[jaxa]{Toshifumi Yanagisawa}
%\author[yyy]{Syohei Miyaji}
%\author[zzz]{Jyunji Suzuki}

\address[nao]{National Astronomical Observatory, Osawa 2--21--1, Mitaka, Tokyo, Japan}
\address[bit]{Department of Astronomy, Bandung Institute of Technology, Jalan Ganesha 10, Bandung 40132, Indonesia}
\address[htd]{Teikyo-Heisei University, 2--51--4 Higashi--Ikebukuro, Toshima, Tokyo 170--8445, Japan}
\address[nro]{Nobeyama Radio Observatory, Nobeyama 462--2, Minami-Maki-Mura, Nagano 384--1305, Japan}
\address[uba]{Ulugh Beg Astronomical Institute, 33 Astronomical Street, Tashkent 700052, Uzbekistan}
\address[lpl]{Lunar \& Planetary Laboratory, The University of Arizona, 1629 E. University Boulevard, Tucson, AZ 85721--0092, USA}
\address[ncu]{Institute of Astronomy, National Central University, Jhongda Rd. 300, Jhongli, Taoyuan 32001, Taiwan}
\address[fue]{Fukuoka University of Education, Akama-Bunkyo-machi 1--1, Munakata, Fukuoka 811--4192, Japan}
% \address[tmo]{Tokyo Metropolitan Government, Nishishinjuku 2--8--1, Shinjuku, Tokyo 163--8001, Japan}
% \address[hkd]{IFES--GCOE Global COE Program, Hokkaido University, Kita 8 Nishi 5, Kita-ku, Sapporo, Hokkaido 060--0810, Japan}

% \address[jaxa]{Japan Aerospace Exploration Agency, Jindaiji-higashi-machi 7--44--1, Chofu, Tokyo, 182--8522, Japan}

\begin{abstract}
The Karin family is a young asteroid family
formed by an asteroid breakup 5.8 Myr ago.
Since the members of this family probably have not experienced
significant orbital or collisional evolution yet,
it is possible that they still preserve properties of
the original family-forming event in terms of their spin state.
We carried out a series of photometric observations of the Karin family
asteroids, and here we report on the analysis of the lightcurves including
the rotation period of eleven members.
%% as well as those of an interloper asteroid:
%  (832) Karin,      % ($P= 18.35 \pm 0.02$),
% (4507) 1990 FV (an interloper), % ($P=  6.58 \pm 0.04$; 
% (7719) 1997 GT36,  % ($P= 29.56 \pm 0.60$),
%(10783) 1999 RB9,   % ($P=  7.33 \pm 0.04$),
%(11728) Einer,      % ($P= 13.62 \pm 0.05$),
%(13765) Nansmith,   % ($P= 10.51 \pm 0.01$),
%(16706) Svojsik,    % ($P=  6.72 \pm 0.07$), 
%(28271) 1999 CK16,  % ($P=  5.64 \pm 0.06$),
%(40917) 1999 TR171, % ($P=  6.74 \pm 0.08$),
%(43032) 1999 VR26,  % ($P= 32.89 \pm 0.04$),
%(69880) 1998 SQ81,  % ($P=  7.68 \pm 0.01$),
%and
%(71031) 1999 XE68.  % ($P= 20.19 \pm 0.41$).
% As for four of them we estimated their absolute magnitudes $H_R$ and
% the slope parameter $G_R$ of the solar phase curves:
%   (832) Karin,      % ($H_R = 11.03 \pm 0.02$, $G_R = 0.19 \pm 0.04$), 
%  (4507) 1990 FV,    % ($H_R = 12.09 \pm 0.02$, $G_R = 0.19 \pm 0.05$), 
% (13765) Nansmith,   % ($H_R = 14.82 \pm 0.02$, $G_R = 0.23 \pm 0.06$),
% and
% (69880) 1998 SQ81.  % ($H_R = 14.60 \pm 0.01$, $G_R = 0.24 \pm 0.05$).
%
The mean rotation rate of the Karin family members turned out to be
much lower than those of near-Earth asteroids or
small main belt asteroids (diameter $D < 12$ km), and
even lower than that of large main belt asteroids ($D>130$ km).
We investigated a correlation between the peak-to-trough variation
% reduced to zero solar phase angle
and the rotation period of the eleven Karin family asteroids, and
found a possible trend that elongated members have lower spin rates,
and less elongated members have higher spin rates.
However, this trend has to be confirmed by another series of future
observations.
\end{abstract}


\end{frontmatter}

% \linenumbers

%%%%%%%%%%%%%%%%%%%%%%%%%%%%%%%%%%%%%%%%%%%%%%%%%%%%%%%%%%%%%%%%%%%%%%%%%%%%%
\section{Introduction\label{sec:intro}}

Asteroid families are remnants of catastrophic disruption and reaccumulation
events between small bodies in the solar system \citep[e.g. ][]{michel2003}.
Each member of
an asteroid family has the potential to provide us with clues about the
family-formation events that created them.
However, since asteroid families are generally old ($\sim$Gyr),
it is quite likely that the family members
have undergone significant orbital, collisional, and
spin-state evolution that masks properties of the original
family-forming events.

A sophisticated numerical technique devised by
\citet{nesvorny2002} changed the above situation.
Using their method, they detected three young 
asteroid families in the main belt:
the Karin family   ($\sim$5.8 Myr old),
the Iannini family ($\sim$5   Myr old), and
the Veritas family ($\sim$8   Myr old).
These families are remarkably younger than previously known asteroid families,
and more and more younger asteroid clusters have been recognized since then
\citep[e.g. ][]{nesvorny2006b,vokrouhlicky2008,vokrouhlicky2009}.
With these discoveries in hand, we find many aspects
of the study of young asteroid families interesting:
their spin period distribution,
their shape distribution, and
possible detection of non-principal axis rotation.

We expect that the young family members preserve some properties of
the original family-forming event in their spin period distribution.
Although there are several laboratory experimental studies 
on the spin period distribution of collisional fragments
\citep[e.g. ][]{fujiwara89,nakamura91,kadono2009},
it is hard to directly apply their results to real collisions
between Small Solar System Bodies (SSSBs) in the gravity-dominant regime.
Thus, observations of spin rates of the young asteroid family members
can be unique opportunities to collect information on large-scale collisions.

As for the asteroid spin rate distribution, it is now widely known that 
the Yarkovsky--O'Keefe--Radzievskii--Paddack (YORP) effect may spin up or
spin down 10-km-sized asteroids on a $10^{8}$ yr timescale, and
smaller asteroids could spin up/down even faster
\citep[e.g. ][]{rubincam2000,bottke2006}.
However, as the ages of the young asteroid families are
substantially shorter than the timescale of the YORP effect, 
each family member perhaps statistically retains its initial
spin status just after the family-formation event.
In old asteroid families, such as the Koronis family, the YORP effect
has changed the initial spin rate since the family-formation events
\citep[e.g. ][]{slivan2002,slivan2003,vokrouhlicky2002,vokrouhlicky2003}.
Comparisons between the spin rate distribution of old and
young asteroid families can serve as a help
in the timescale estimate of the YORP effect.
%%% Added acording to Referee 1's order
Actually the YORP effect is very sensitive to small-scale
topography of asteroids \citep[e.g. ][]{statler2009}.
However, with the current observational data that we have in hand,
we have not reached a detailed quantitative estimate of how seriously
the YORP effect has influenced the dynamics of the Karin family members.
Gaining a deeper understanding of these dynamics remains an aim of
future inquiry.

In addition to the spin rate statistics,
the shape distribution of the young asteroid family members is
important for understanding the fragmentation and reaccumulation process of
SSSBs in comparison with laboratory
collisional experiments. 
It may help us understand the dynamical process of fragmentation and
reaccumulation of asteroids, such as how angular momentum is distributed
to each of the remnants.
Also, it is possible to get an estimate of the satellite/binary
forming efficiency at asteroid disruption events.

\comment{
Third,
surface color of the young family asteroids is quite
interesting in its relation to the space weathering process, as
it is now well known that the spectra of asteroid surface change through 
space weathering \citep[e.g. ][]{chapman96,chapman2004,mcfadden2001,clark2002}.
Detailed physics and mechanism of the space weathering process have been
partially reproduced by laboratory experiments
\citep[e.g. ][]{sasaki2001,marchi2005,hiroi2006}.
Spectroscopic observations of the young family asteroids have been
carried out \citep[e.g. ][]{mothe-diniz-nesvorny2008},
and particularly as for the largest member of the Karin family,
(832) Karin, an arguable discussion as to whether or not this asteroid
has an inhomogeneous surface color is going on
\citep{sasaki2004,yoshida2004a,chapman2007,vernazza2007,ito2007a}.
A special feature that the young asteroid families have is that we can
use them as a precise measure for making a quantitative relationship between
the asteroid surface age and the degree of space weathering
\citep[e.g. ][]{jedicke2004,nesvorny2005,vernazza2009,willman2010}.
}

% In the near-Earth asteroids (NEAs) population,
% \citet{binzel96} found that several asteroids show intermediate
% spectra between the S-type asteroids and ordinary 
% chondrites. It has been interpreted that the intermediate spectra were
% produced by the space weathering process that changes optical properties
% of asteroid surfaces

The young asteroid families also draw our
attention in terms of possible detection of non-principal axis rotation
(sometimes called ``tumbling motion'').
The study of a celestial body's non-principal axis rotation gives us important
insights into energy dissipation and excitation processes,
as well as internal structure of the body.
Non-principal axis rotation could be excited by collisions of small projectiles,
but it will be damped quickly unless the excitation continues.
This is the main reason why
the non-principal axis rotation of SSSBs has been
confirmed only for a few tens of lightcurves
\citep[e.g. ][]{harris94,pravec2000,paolicchi2002,mueller2002,warner2009,oey2012,pravec2014}.
However, the age of the young family asteroids is quite young,
and we may be able to observe their non-principal axis rotation before
it has totally decayed.

Based on the motivations mentioned above, we began a series of 
photometric observations of the young asteroid families
in November 2002.
In this paper we focus on the current result of our lightcurve observation of
the Karin family asteroids through the $R$-band imaging that we had carried
out until May 2004, and summarize the result for eleven Karin family
members whose rotation period we determined.
Note that throughout the present paper we assume that the lightcurve
variations are due to shapes of asteroids, not due to albedo features.

% Section \ref{sec:obs} describes details of our observation.
% Section \ref{sec:lc} summarizes
%   the method of lightcurve analysis (\ref{ssec:lc-method}) and
%   resulting lightcurves (\ref{ssec:lc-lc}).
% Section \ref{sec:summary} is devoted to discussions on the results.

%%%%%%%%%%%%%%%%%%%%%%%%%%%%%%%%%%%%%%%%%%%%%%%%%%%%%%%%%%%%%%%%%%%%%%%%%%%%%%
\section{Observations\label{sec:obs}}

During the period from November 2002 to May 2004,
we observed and determined the rotational periods of eleven
Karin family members, including the largest member, (832) Karin.
\comment{
  (832) Karin,
 (7719) 1997 GT36,
(10783) 1991 RB9,
(11728) Einer,
(13765) Nansmith, 
(16706) Svojsik,
(28271) 1999 CK16,
(40912) 1999 TR171,
(43032) 1999 VR26,
(69880) 1998 SQ81,
and
(71031) 1999 XE68.
We used the following telescopes for our observations:
the 2.3 m telescope (``Bok'') at the Steward Observatory
          (Kitt Peak, Arizona, USA),
the 1.8 m telescope (``VATT'') at the Vatican Observatory
          (Mt. Graham, Arizona, USA),
the 1.5 m telescope (``AZT'') at Maidanak Observatory (Uzbekistan), 
the 1 m telescope at the Lulin Observatory (Taiwan)
the 1 m telescope at the Kiso Observatory (Nagano, Japan),
the 0.5 m telescope at the National Astronomical Observatory (Tokyo, Japan),
the 0.4 m telescope at the Fukuoka University of Education (Fukuoka, Japan),
and
the 0.25 m telescope at the Miyasaka Observatory (Kobuchizawa, Japan).
% and
% the 6.5 m telescope (``Magellan'') at the Las Campanas Observatory (Chile).
}
Table \ref{tbl:table1} shows the list of the observatories, the telescopes,
and field of views of the instruments that we used for our observations.
%%%%%
\marginpar{[Table \ref{tbl:table1}]}
%%%%%

We used the $R$-filter for our lightcurve observations because
it is widely known that brightness of the reflected light in
optical wavelengths from most asteroids becomes the highest in the $R$-band
among the Johnson--Cousins $UBVRI$ filters.
In our observations all the telescopes were driven
at the sidereal tracking rate, and the exposure time was limited by the
moving rate of asteroids as well as by seeing during the
observing nights. As typical main belt asteroids (MBAs)
having the semimajor axis $a=2.8$ AU
move at the speed of $\sim 0.55''$/min at its opposition, and as
the typical seeing size at the observatories was 
from $1.0''$ to $3.0''$,
we chose a single exposure time of two to eight minutes so that
an asteroid has an appearance of a point source.
% In principle,
Generally,
we continued the $R$-band imaging for a particular asteroid
throughout a night except when we took images of standard stars:
an ``asteroid per night'' strategy.

We used the Landolt standard stars \citep{landolt92}
for the purpose of calibration.
% Every night we observed several Landolt photometric standard stars
% \citep{landolt92} at several airmasses to determine extinction
% coefficients
% (note that we used the standard stars just for this purpose,
%  not for calibrating zero-points to the standard system).
%%%
% See Section \ref{subsec:combineLCs} for detail about our standard star observation and calibration.
Before and/or after each of the observing nights, we took
dome flats or twilight sky flats for flat-fielding.
After the observation, we applied a standard data reduction procedure
against the data: bias subtraction and flat division.
% First, we subtracted bias images from object images
% and from flat images.
% Then, each of the images was divided by the median flat-fielding image
% that was created by calculating the median of all the flat images.
% Then we carried out the aperture photometry using the apphot package in IRAF.
%
% Brightness of an asteroid was measured relative to that of a field star
% located on the same frame as the asteroid. We chose the field stars
% from the USNO--A2.0 star catalogue%
% \footnote{%
% \sf http://tdc-www.harvard.edu/catalogs/ua2.html
% }.
% We corrected the magnitude of asteroids
% using the extinction curve obtained on each of the observing nights,
% using the standard star observation.
Table \ref{tbl:table2} is the summary of our observational details.
%%%%%
\marginpar{[Table \ref{tbl:table2}]}
%%%%%

%%%%%%%%%%%%%%%%%%%%%%%%%%%%%%%%%%%%%%%%%%%%%%%%%%%%%%%%%%%%%%%%%%%%%%%%%%%%
\section{Analysis and results\label{sec:lc}}

\comment{
Since we use different telescopes with different instruments and we
combine lightcurves obtained at different date and at different
aspects, we need careful calibrations for making a lightcurve of
an asteroid obtained from multiple observing run.

At first, we
estimate a rough rotational period of asteroid by using the data
obtained from consecutive observing nights. Here we employed two
different algorithms to examine a periodicity in the lightcurve data:
Lomb's Spectral Analysis \citep[LSA, ][]{lomb76},
and the WindowCLEAN Analysis \citep[WCA, ][]{roberts87}.
WCA incorporates a discrete Fourier transform and the CLEAN algorithm
\citep{hogbom74}, and it was used for detection of multiple periods
on asteroid (4179) Toutatis \citep{mueller2002}.
Using the two different method,
when the highest spectral powers of spin periods obtained by
by both the analyses are consistent to each other, the corresponding
period was used as the ``true'' spin period of the asteroid.
When the spectrum powers obtained by each of the methods differ from
each other, 
\textcolor{red}{??????????}.
}

To construct composite lightcurves of asteroids from the
observational data, we followed a sequence proposed by \citet{harris89}.
The actual procedure is described in our previous publications
\citep{dermawan2002,dermawan2011,yoshida2004a}.
Principally, it is an iterative repetition of frequency analysis
and fitting to Fourier series.
We employed two different algorithms to examine periodicities in
the lightcurve data: Lomb's Spectral Analysis \citep[LSA, ][]{lomb76} and
the WindowCLEAN Analysis \citep[WCA, ][]{roberts87}.
WCA incorporates a discrete Fourier transform as well as the CLEAN algorithm
\citep{hogbom74}, and \citet{mueller2002} adopted WCA when they detected
multiple rotational periodicities of asteroid (4179) Toutatis.
When the frequency analysis is done, we fit the lightcurve
with a Fourier series.
We have to be particularly careful when we combine lightcurves
derived from several observing runs because they generally have different
lightcurve-mean magnitudes.
% (i.e. mean magnitude of the fitting curve by Fourier series).
See Section \ref{subsec:combineLCs} for details of how we combined
the lightcurves obtained from multiple observing runs.

% End of \comment
%%%%%%%%%%%%%%%%%%%%%%%% Comment out up to here %%%%%%%%%%%%%%%%%%%%%%%%%%%%

Once we have obtained the lightcurve of an asteroid,
we estimate the peak-to-trough variation of its lightcurve.
%
% Incorporating Reviewer 1's suggestion
To compare the amplitudes $(A)$ of the lightcurves of the Karin
family members taken at different solar phase angles $(\alpha)$
with each other as well as with other solar system bodies,
we used the empirical relationship by \citet{zappala90}
that normalizes the amplitudes to a solar phase angle of 0 degree.
\citet{zappala90} gives
$ A(\alpha) = A(0) \left( 1+m\alpha \right) $,
and it empirically determines the parameter $m = 0.030$ for S-type asteroids,
which the Karin family members are classified as.
However, we have to note that these amplitudes can be only used in
a statistical sense, because, except for (832) Karin,
these asteroids' spin obliquities are not known.

%%%%%%%%%%%%%%%%%%%%%%%%%%%%%%%%%%%%%%%%%%%%%%%%%%%%%%%%%%%%%%%%%%%%%%%%%%%%%%
%%% The former Appendix A
%%%%%%%%%%%%%%%%%%%%%%%%%%%%%%%%%%%%%%%%%%%%%%%%%%%%%%%%%%%%%%%%%%%%%%%%%%%%%%
\subsection{Procedure for combining lightcurves\label{subsec:combineLCs}}
In this subsection
we describe how we dealt with the standard stars in our observation
and how we combined lightcurves of asteroids obtained from different
observing nights, making a single lightcurve for each asteroid.

\subsubsection{Observation of the Landolt standard stars\label{subsec:A_1}}
We took the following procedures when observing Landolt standard stars.

% \label{subsubsec:A_1_1}
\paragraph{(1) On photometric nights}
Before and/or after the observation of each asteroid,
we take images of the Landolt standard stars at several different airmasses.
We determine the atmospheric extinction coefficients of the night
based on this dataset.

% \label{subsubsec:A_1_2}
\paragraph{(2) On non-photometric nights}
While we take images of asteroids,
we take images of one or two Landolt standard stars,
once or twice if possible,
at the same airmasses at which we observed the asteroids.
% Later, we use the data from this observation in the process, as described in \ref{subsec:A_3}.

\subsubsection{Combining lightcurves I. (with clear maxima and minima)\label{subsec:A_2}}
When all the observations are done, we measure the brightness of
asteroids by IRAF. Eventually we combine all the lightcurves from
different observing nights into a single composite lightcurve for
each asteroid.

% \label{subsubsec:A_2_1}
\paragraph{(1) Measuring field stars}
When we measure the brightness of an asteroid by IRAF,
we also measure the brightness of five or six field stars
that exist on the same image as the asteroid.
Here we should note that all the field stars that we choose must be included
in the USNO--A2.0 catalogue%
\footnote{%
\sf http://tdc-www.harvard.edu/catalogs/ua2.html
}.
Although we are aware that the brightness magnitude of the stars
catalogued in the USNO--A2.0 catalog contains a certain degree of error,
here we do not take care of the errors.

% \label{subsubsec:A_2_2}
\paragraph{(2) Elimination of anomalous field stars}
We draw lightcurves of each of the field stars, and confirm
that none of those we have chosen is a variable star.
Also, if any of the chosen field stars is located on an image frame
with a serious problem such as caused by camera shutter problems or
cosmic ray contamination,
we discard the entire image itself from the analysis.

% \label{subsubsec:A_2_3}
\paragraph{(3) Choosing the brightest field star}
When we have found several field stars whose brightness does not vary
with time, we select the brightest one among them as a comparison star
for the asteroid's relative photometry. This is because errors in
relative photometry are smaller when we choose a brighter star
as a comparison star.

% \label{subsubsec:A_2_4}
\paragraph{(4) Relative photometry of asteroids}
Using the selected field stars as comparison stars, we carry out
relative photometry work for each asteroid.
As a result, we obtain lightcurves in relative magnitude
for each of the observation nights for each of the asteroids.

% \label{subsubsec:A_2_5}
\paragraph{(5) Determination of lightcurve-mean magnitudes}
We determine lightcurve-mean magnitude of each asteroid from the
lightcurves in relative magnitude.
We get the lightcurve-mean magnitudes in the process of fitting each of the
lightcurves with Fourier series.
When we see maximum and minimum in lightcurves, determination of
lightcurve-mean magnitude is straightforward.

% \label{subsubsec:A_2_6}
\paragraph{(6) Combining lightcurves based on the lightcurve means}
When several observing nights are (nearly) consecutively distributed,
we combine the lightcurves of each asteroid from all the observing nights,
based on the lightcurve-mean magnitude that we have determined
in \ref{subsec:A_2} (5).

% \label{subsubsec:A_2_7}
\paragraph{(7) Frequency analysis}
For each asteroid, we carry out frequency analysis of the
combined lightcurves produced in \ref{subsec:A_2} (6)
and get necessary quantities such as rotation period.

\subsubsection{Combining lightcurves II. (without clear maxima or minima)\label{subsec:A_3}}
When we do not see any maxima or minima in lightcurves of asteroids,
determination of lightcurve-mean magnitude is not straightforward.
This happens when the spin period of an asteroid is as long as,
or longer than, the observational period---observation length of a
night does not reach the spin period of an asteroid.
In this case we utilize the observational data of the Landolt standard stars
that we have prepared in
  \ref{subsec:A_1} (1)   and \ref{subsec:A_1} (2),
% \ref{subsubsec:A_1_1} and \ref{subsubsec:A_1_2},
and carry out the following procedures, instead of
  \ref{subsec:A_2} (5)   and \ref{subsec:A_2} (6),
% \ref{subsubsec:A_2_5} and \ref{subsubsec:A_2_6},
to combine several lightcurves of an asteroid into a single one.
Note that we carry out the procedure described in this subsection
only when the observation night is photometric.
Observational data from non-photometric nights without clear maxima or minima
is not used, and is just discarded.

% \label{subsubsec:A_3_1}}
\paragraph{(1) Calibration of the field star brightness}
% \paragraph{(a) Data from photometric nights}
Using the Landolt standard stars, we calibrate the brightness of the
field stars that we selected. Then we calculate the apparent brightness
magnitude of the asteroid using the calibrated brightness of the field stars.

\comment{
\paragraph{(b) Data from non-photometric nights}
Using the extinction coefficients obtained from the data at the photometric
nights (see \ref{subsubsec:A_1_1}),
we estimate the magnitude of the Landolt standard stars
whose magnitude we obtained at the same airmasses with the asteroids
and the field stars that we selected. Strictly speaking, 
extinction coefficients of the Landolt standard stars must be different
between at photometric and at non-photometric nights, but here we assume
their difference is negligible.
Next we derive the cataloged magnitudes of the selected field stars
from the USNO--A2.0 catalog, and calibrate them using the estimated
magnitude of the Landolt standard stars that we mentioned above.
Finally, we calibrate the asteroids' magnitude using the calibrated
magnitude of the field stars,

We are aware that the procedure mentioned here is based on an uncertain
assumption, but it is only what we can do for the data in this case, and
the amount of data that fall in this category is not much (there are
only four NP${}^\ast$s in Table \ref{tbl:table1}).
So we think the addition of uncertainty by this assumption in the entire
lightcurve analysis is limited, although it may have a non-negligible
influence on the lightcurve analysis of (16706) Svojsik where half of
the lightcurve data fall in the category of NP${}^\ast$s.
}

% \label{subsubsec:A_3_2}
\paragraph{(2) Combining lightcurves based on the lightcurve means}
Based on the magnitude of each of the asteroids estimated in
  \ref{subsec:A_3} (1),
% \ref{subsubsec:A_3_1},
we combine their lightcurves from different nights into a single lightcurve.

% \label{subsubsec:A_3_3}
\paragraph{(3) Frequency analysis}
For each asteroid, we carry out frequency analysis of the
combined lightcurves produced in
  \ref{subsec:A_3} (2),
% \ref{subsubsec:A_3_2},
and get necessary quantities such as rotation period.

\subsection{Lightcurves\label{ssec:lc-lc}}
%%%%%
\marginpar{[Fig. \ref{fig:lc_all}]}
\marginpar{[Table \ref{tbl:table3}]}
%%%%%

Fig. \ref{fig:lc_all} shows all the lightcurves that we obtained
in the series of observations.
Table \ref{tbl:table3} summarizes
rotation period $P$, reduced peak-to-trough variation $A(0)$,
solar phase angle $\alpha$ during the observation period, and
the period quality code \citep{lagerkvist89}.
As for the rotation period $P$, we chose the most reliable peak value
from the periodicity analysis results by LSA or WCA
(mostly by WCA for the lightcurves presented in the present paper).
%
% Corrected the confusion that we caused on Reviewer 1's understanding
%
We regard that the LSA analysis as compensating for vulnerabilities of WCA
and justifying our result even when the lightcurve data contains large
temporal gaps. 
%
% Added along with Reviewer 1's suggestion
%
Note that we checked out phase plots for the peaks of other periods,
although we did not show them in the present paper.

Since Fig. \ref{fig:lc_all} and Tables \ref{tbl:table2} and \ref{tbl:table3}
describe most of our results, we just give supplementary information
for three of the objects as follows:

\paragraph{(832) Karin} %
The results of our lightcurve observation of this asteroid are
already published (\citet{yoshida2004a} for the observation in 2003,
and \citet{ito2007a} for the observation in 2004).
Since our 2004 observation was mainly for multi-color photometry of
this asteroid, here we just present our 2003 observation result
from \citet{yoshida2004a}.
% Note that we obtained the same spin period, $P = 18.35 \pm 0.02$ hours,
% from both our observations in 2003 and 2004.
%% This is largely consistent with the value that \citet{binzel87}
%% obtained in their 1984 observation, $P = 18.82 \pm 0.10$ hours.

\paragraph{(28271) 1999 CK${}_{16}$} %
We observed this asteroid twice at two different oppositions:
from November to December 2002 and in March 2004.
The rotation periods that were derived from both the observations
are close to each other.
The lightcurve amplitudes taken in 2002 (Fig. \ref{fig:lc_all}(g))
and 2004 (Fig. \ref{fig:lc_all}(h)) are different because geometric
configurations between the asteroid, observer, and the Sun were different.
%%% Removed by the suggestion of Reviewer 2
% If we observe this asteroid at another configuration we will
% be able to obtain the spin vector direction, and hopefully
% a shape model of this asteroid.

\paragraph{(11728) Einer} %
The rotation period derived for this asteroid is from the most prominent
peak obtained from the period analysis.
However, note that other period values might also be possible due to
potential aliasing.

\noindent
\noindent
% $\hrulefill$

%%%%%%%%%%%%%%%%%%%%%%%%%%%%%%%%%%%%%%%%%%%%%%%%%%%%%%%%%%%%%%%%%%%%%%%%%%%
\section{Discussions\label{sec:summary}}

%%%
%%% On the average spin periods
%%%
The spin period distribution of asteroids is often compared with
the Maxwellian distribution \citep[e.g. ][]{binzel89}.
Unfortunately, the number of our lightcurve samples is still
far from being sufficient for such a detailed statistical discussion.
Here, let us just compare the mean value of the rotation rate $1/P$
of the eleven Karin family asteroids that we observed
with those of near-Earth asteroids (NEAs),
small MBAs ($D<12$ km), and large MBAs ($D>130$ km).
According to Table~2 of \citet[][p.~265]{binzel2002}, the mean values of
the rotation rate of NEAs, small MBAs, and large MBAs are
% $5.00^{5.32}_{4.72}$ hours,
% $5.53^{5.84}_{5.25}$ hours, and
% $8.28^{8.63}_{7.95}$ hours, respectively.
$4.80 \pm 0.29$ rev/day,
$4.34 \pm 0.23$ rev/day, and
$2.90 \pm 0.12$ rev/day, respectively.
On the other hand, from our present work,
the mean rotation rate of the Karin family asteroids turned out to be
%%% $10.17^{20.69}_{6.74}$ hours
% $\sim$0.93 rev/day, or
% $\sim$0.98 rev/day excluding (832) Karin.
%
% (20150302) The above numbers are wrong! To check them out, try:
% grep -v ^# fig2data.dat | awk '{printf ``%f %f\n'', $2, 24/$2}' | awk -f ~/bin/sum.awk
% The correct values are as follows with a new footnote:
%%% BEFORE excluding NP* data
% $\sim$2.36 rev/day, or
% $\sim$2.46 rev/day excluding (832) Karin
%%% AFTER excluding NP* data
$\sim$2.40 rev/day, or
$\sim$2.51 rev/day excluding (832) Karin
(we used the period value obtained in 2004 for (28271) 1999CK${}_{16}$).
Therefore,
the mean rotation rate of the Karin family members is much lower
than those of the NEAs and the small MBAs, and even lower than
that of the large MBAs.
This may be quite an interesting fact, considering the widely believed
hypothesis that most of the small MBAs are collisional remnants.

According to Table~2 of \citet[][p.~265]{binzel2002},
the reduced peak-to-trough variation of
the NEAs,
the small MBAs, and
the large  MBAs
is
0.29,
0.28, and
0.19, respectively.
Meanwhile, the average reduced peak-to-trough 
variation of the Karin family members is
%%% BEFORE excluding NP* data
%0.23--0.27%
%%% AFTER excluding NP* data
0.24--0.27%
\footnote{It is
%%% BEFORE excluding NP* data
%$\sim$0.26 if we choose the value 0.06 for (28721) 1999 CK${}_{16}$,
%$\sim$0.23 if we exclude (832) Karin and choose the value 0.06 for (28721) 1999 CK${}_{16}$,
%$\sim$0.27 if we choose the value 0.17 for (28721) 1999 CK${}_{16}$, and
%$\sim$0.24 if we exclude (832) Karin and choose the value 0.17 for (28721) 1999 CK${}_{16}$.
%%% AFTER excluding NP* data
$\sim$0.26 if we choose the value 0.07 for (28721) 1999 CK${}_{16}$,
$\sim$0.24 if we exclude (832) Karin and choose the value 0.07 for (28721) 1999 CK${}_{16}$,
$\sim$0.27 if we choose the value 0.17 for (28721) 1999 CK${}_{16}$, and
$\sim$0.25 if we exclude (832) Karin and choose the value 0.17 for (28721) 1999 CK${}_{16}$.
}%
.
The average value excluding (832) Karin, 
%%% BEFORE excluding NP* data
% 0.23--0.24,
%%% AFTER excluding NP* data
0.24--0.25,
is closer to that of the small MBA group with $D<12$ km,
rather than that of the large MBAs.
This is consistent with our conventional knowledge that
asteroid remnants, such as small MBAs or young family asteroids,
are more likely to have an elongated (and irregular) shape than a
spherical shape compared with large asteroids that can be
parent bodies of asteroid families.

%%%
%%% On the ``trend'' conjecture
%%%
%%%%%
\marginpar{[Fig. \ref{fig:stat}]}
%%%%%
We summarized 
our main result of the spin period $P$ and the peak-to-trough variation
magnitude $A(0)$ of the eleven Karin family members in Fig. \ref{fig:stat}
(we are aware that we have ignored the effect of asteroids'
obliquity in this figure, because we have no information on it so far).
In Fig. \ref{fig:stat}(a) that shows the relation between $A(0)$ and $1/P$,
you may see a slight trend from the top left to the bottom right,
which tells us that elongated asteroids have a lower spin rate,
and those less elongated asteroids have a higher spin rate.
A similar trend has been recognized in fast-rotating sub-km-size
MBAs \citep{dermawan2011,nakamura2011}.
% If the rotational angular momentum given to each remnant is nearly
% equally distributed when an asteroid family is formed,
% it is possible that the smaller the remnant is, the faster its rotation
% becomes (which is already confirmed by laboratory experiments
% such as \citet{kadono2009}).
However, we are aware that the number of our present lightcurve samples is
not large enough to reach a definite conclusion on our conjecture.
%%% Reviewer 1's opinion
In other words, Fig. \ref{fig:stat}(a) may be just a scatter plot
from the fact that the number of objects is not large enough
for rigorously statistical discussions.
Whether the ``trend'' really exists depends on how many more lightcurve
samples of the Karin family members we can obtain from now on.
% Also, the size range of our observing targets is somehow limited
% within $D$=3--5 km except (832) Karin (Fig. \ref{fig:stat}(b)(c)),
% possibly causing the appearance of the ``trend''.

% Also, if the mass of the remnants are
% not so different from each other, it is also possible that
% the more spherical the remnant is, the faster its rotation becomes.
% This is actually the case for the Karin family members that are
% considered here, as seen in Fig. \ref{fig:stat}(c) and
% Fig. \ref{fig:stat}(d):
% Most of the family members except the largest one, (832) Karin, are
% concentrated in a narrow size range between $3 \lesssim D \lesssim 5$ km.
% Hence, the trend seen in Fig. \ref{fig:stat}(a) might be a result of
% equi-distribution of the rotational angular momentum when this family
% was created 5.8 Myr ago.

%%%
%%% On the non-principal axis rotation
%%%
%%%%%
\marginpar{[Fig. \ref{fig:D-P}]}
%%%%%
At the end of this paper, we would like to consider the possibility that
some of the Karin family members still possess non-principal axis rotation.
From the result that we have presented in the previous sections,
we plotted the rotation period of the eleven Karin family asteroids on a
diagram that shows the relation between rotation period $P$ and diameter $D$
in Fig. \ref{fig:D-P}.
For this figure, we estimated the diameter of the Karin family members
from its absolute magnitude $H$ using the relationship
$ \log_{10} D = 3.1295 - 0.5 \log_{10} p - 0.2 H $
where $p$ is the albedo of an asteroid \citep{bowell79}.
%
% D-H relation
%
%\begin{equation}
% \log_{10} D = 3.1295 - 0.5 \log_{10} p - 0.2 H,
%  \label{eqn:D-H}
%\end{equation}
% As for (832) Karin, (13765) Nansmith, and (69880) 1998SQ${}_{81}$, we used
% their absolute magnitude obtained from our own observations
% (Table \ref{tbl:table3}).
%%% \textcolor{red}{\bf (Q: What about 1990 FV? We have its $H_R$ value.)}
% For other members that we have observed, 
We used the absolute magnitude values of the asteroids
listed on the Lowell asteroid orbital elements database%
%%%MPC database
\footnote{%
%\sf http://www.minorplanetcenter.net/iau/MPCORB.html
\sf ftp://ftp.lowell.edu/pub/elgb/astorb.html
}.
We assumed the albedo value of $p=0.21$ for the S-type asteroids
\citep[cf. ][]{yoshida2007,strom2015}%
\footnote{%
The value $p=0.21$ is the mean albedo calculated from the catalog of
asteroid albedo and taxonomic types in PSI PDS database,
{\sf http://sbn.psi.edu/pds/resource/albedo.html}.
The albedo value can be dependent on asteroid size, but we are not
sure how strongly or weakly dependent it is. For example, recent
survey observations have revealed the albedo dependence on
asteroid size for each of the taxonomic types 
\citep[e.g. ][]{usui2013}. 
In \citeauthor{usui2013}'s (\citeyear{usui2013}) Figure 5(b)
for S-type asteroids, we may (or may not) see some skewed albedo
distribution in the diameter range of $D < 10$km, but we are not sure
about the quantitative analysis result in a smaller range.
Another thing is that the range of $D$ of the asteroids that we
observed this time is rather narrow (see Figure \ref{fig:stat}(c))
except (832) Karin, and we can expect that the albedos for these asteroids
are not so different from each other, even if there is a dependency of
albedo on size. Therefore, in the present paper, we used the average albedo.
}.

For comparison, we also plotted the $(D,P)$ relation of
3,745 known asteroids listed on the PSI PDS lightcurve database
whose rotational periods are known to a certain reliability%
\footnote{%
{\sf http://sbn.psi.edu/pds/resource/lc.html} as of April 30, 2012 (V12.0).
Among the lightcurve datafile {\sf data/lc\_{}summary.tab},
we selected the asteroids only with the lightcurve reliability of
$2, 2+, 3$ or higher.
}.
For these asteroids, we applied the mean albedo of $p=0.081$ following
\citet{ryan2010} to all the asteroids, assuming their absolute magnitude
$H$ listed in the Lowell asteroid orbital elements database mentioned above.
Also, among these asteroids we highlighted 31 possible tumblers%
\footnote{%
Among all the lightcurve data in {\sf data/lc\_{}summary.tab},
we selected the asteroids having the credible tumbling flag
{\sf T} or {\sf T+}.
}
in green so that we can compare their $(D,P)$ relation with that of the 
Karin family asteroids.

Theoretically, the damping timescale of the non-principal axis rotation
of a celestial body $T_{\rm d}$ (Gyr)
can be expressed in its relationship between $P$ (hours) and $D$ (km)
as in the following equation by \citet{harris94},
a reconsideration of a theory by \citet{burns73}:
%
% D,P,Td relation
%
\begin{equation}
% T_{\rm d} \sim 0.2 \frac{P^3}{D^2},
% P^3 \sim 4.9 D^2 T_{\rm d},
  P \sim 17 D^\frac{2}{3} T_{\rm d}^\frac{1}{3} .
  \label{eqn:damping}
\end{equation}
Smaller and slower rotators have
longer damping timescales of the non-principal axis rotation.
In Fig. \ref{fig:D-P} we drew two damping timescales of non-principal axis rotation
using diagonal lines calculated by Eq. (\ref{eqn:damping}).
The upper solid blue line in Fig. \ref{fig:D-P} indicates the
$(D,P)$ relation of asteroids when their damping timescale
$T_{\rm d} = 5.8$ Myr,
equivalent to the age of the Karin family.
The lower dashed blue line in Fig. \ref{fig:D-P} indicates
that of asteroids when their damping timescale $T_{\rm d} = 4.6$ Gyr,
almost equivalent to the age of the solar system.
You can see from Fig. \ref{fig:D-P} that many of the Karin members
we observed are located below the upper solid line for $T_{\rm d} = 5.8$ Myr, 
indicating that they can still maintain the non-principal axis rotation,
if there is any,
since their damping timescale $T_{\rm d}$ is possibly
longer than their age, 5.8 Myr.
Although in the present analysis we did not detect clues on the
non-principal axis rotation of the Karin family members,
several members with relatively long rotational periods such as
 (7719) 1997 GT${}_{36}$,
(43032) 1999 VR${}_{26}$, or
(71031) 1999 XE${}_{68}$
are still candidates as tumbling asteroids.

%%%
%%% Final remark
%%%
% So far, a large part of our observations presented in this paper
% has been carried out only at one opposition per asteroid due to the
% limitation of the telescope resource that were assigned to us
% for this purpose.
% In general, lightcurve amplitude of a particular asteroid can be
% significantly different depending on which opposition or aspect
% you observe the asteroid from. In this sense, a statistics such as in
% Fig. \ref{fig:stat} can be altered or revised
% in future along with the accumulation
% of lightcurve data that our project acquires.

\newpage
%%%%%%%%%%%
%%%% acknowledgments
%%%%%%%%%%%
We greatly appreciate the effort and courtesy extended by all the people
who helped us during our observations at the Vatican Observatory's
 ``VATT'' (the Alice P. Lennon Telescope and the Thomas J.
Bannan Astrophysics Facility).
We thank Elizabeth Green, Paula White, Ed Olszewski, and Andy Odell
for allowing our use of the Steward 90-inch ``Bok'' telescope.
FY is deeply thankful to Tom Gehrels who recommended that we use
the VATT and the Bok telescopes in Arizona, and particularly for
his encouragement of our observing activity and his enormous contribution
to asteroid studies. Without his warmth, help, and thoughtful suggestions,
we could not have carried out our observations there at all.
The staff members at the Maidanak Observatory greatly helped us
during our stay and observing activity.
% We also thank Matthew Holman for providing us with lightcurve data of
% some Karin family asteroids that will be published in future.
% asteroid (51068) 2000 GW156 at the Magellan telescope.
The authors thank the two anonymous referees for suggesting directions
that significantly improved the quality of this paper.
The authors have also benefited from stimulating enlightenment by Murier Yoshida.
Detailed and constructive review by Yolande McLean has
considerably improved the English presentation of this paper.
Some part of the data analysis was performed 
at the National Astronomical Observatory of Japan's
  Astronomy Data Center (ADC) and
   Center for Computational Astrophysics (CfCA).
This study is supported by the JSPS Kakenhi grant
(16740259/2004--2005, 18540426/2006--2008, 18540427/2006--2008,
 21540442/2009--2011, 25400458/2013--2015, 25400238/2013--2015),
the JSPS program for Asia--Africa academic platform (2009--2011),
the JSPS bilateral open partnership joint research project (2014--2015),
the Sumitomo Foundation research funding (030755/2003--2004),
and
the Heiwa Nakajima Foundation research funding for Asian studies (2009).

\clearpage

\bibliographystyle{model2-names}
% \bibliography{mybib}
% \input{bbl_fixed.tex}

%% not want to use model2-names.bst.



%% \bibitem must have one of the following forms:
%%   \bibitem[Jones et al.(1990)]{key}...
%%   \bibitem[Jones et al.(1990)Jones, Baker, and Williams]{key}...
%%   \bibitem[Jones et al., 1990]{key}...
%%   \bibitem[\protect\citeauthoryear{Jones, Baker, and Williams}{Jones
%%       et al.}{1990}]{key}...
%%   \bibitem[\protect\citeauthoryear{Jones et al.}{1990}]{key}...
%%   \bibitem[\protect\astroncite{Jones et al.}{1990}]{key}...
%%   \bibitem[\protect\citename{Jones et al., }1990]{key}...
%%   \harvarditem[Jones et al.]{Jones, Baker, and Williams}{1990}{key}...
%%


% \end{thebibliography}

% \clearpage

%%%%%%%%%%%
%%%% tables
%%%%%%%%%%%
%\section*{Tables}

%%%%table1
%{\large Table1}
%\bigskip

\begin{table}[htbp]
\caption{Observatories and instruments.
$E$ is the elevation of the observatory (m),
$D_{\rm t}$ is the diameter of the telescope mirror that we used (m), and
FOV denotes the field of view of the imaging system that we used for our purpose.
The full observatory names and the telescope names are as follows:
Steward:  the 2.3 m telescope (``Bok'') at the Steward Observatory (Kitt Peak, Arizona, USA).
Vatican:  the 1.8 m telescope (``VATT'') at the Vatican Observatory (Mt. Graham, Arizona, USA).
Maidanak: the 1.5 m telescope (``AZT'') at Maidanak Observatory (Uzbekistan).
Lulin:    the 1 m telescope at the Lulin Observatory (Taiwan).
Kiso:     the 1 m telescope at the Kiso Observatory (Nagano, Japan).
%Tokyo:   the 0.5 m telescope at the National Astronomical Observatory (NAOJ headquarters in Tokyo, Japan).
Fukuoka:  the 0.4 m telescope at the Fukuoka University of Education (Fukuoka, Japan).
%Miyasaka: the 0.25 m telescope at the Miyasaka Observatory (Kobuchizawa, Japan).
% and
% Las Campanas: the 6.5 m telescope (``Magellan'') at the Las Campanas Observatory.
}
%%% for Yolande
%\comment{
%%% for Yolande
\begin{center}
\begin{tabular}{cllrlc}
\hline
Name & Longitude & Latitude & \multicolumn{1}{c}{$E$} & \multicolumn{1}{c}{$D_{\rm t}$} & FOV \\
\hline
Steward     & $111^{\circ}36'01.6''$W   & $31^{\circ}57'46.5''$N   & 2071 & 2.29 & $4.5'\times4.5'$    \\ 
Vatican      & $109^{\circ}53'31.25''$W & $32^{\circ}42'04.69''$N & 3191 & 1.8   &  $6.8'\times6.8'$   \\ 
Maidanak  & $66^{\circ}53'47.08''$E    & $38^{\circ}40'23.95''$N & 2593 & 1.5   &  $8.5'\times3.5'$    \\
Lulin           & $120^{\circ}52'25''$E       & $23^{\circ}28'07''$N       & 2862 & 1.0   &  $11.5'\times11.2'$\\
Kiso            & $137^{\circ}37'42.2''$E    & $35^{\circ}47'38.7''$N   & 1130 & 1.05 &  $50'\times50'$      \\
%Tokyo           & $139^{\circ}32'31.6''$E   & $35^{\circ}40'20.2''$N    & 56.4  & 0.50  &  $15'\times10'$      \\
Fukuoka & $130^{\circ}35'44.7''$E   & $33^{\circ}48'45.3''$N    & 70      & 0.40 &  $5.75'\times4.36'$ \\
%Miyasaka  & $138^{\circ}18'01.1''$E    & $35^{\circ}51'45.0''$N    & 880   & 0.25 &  $19.78'\times14.97'$ \\
%%% Las Campanas & $70^{\circ}42'07.2''$W      & $29^{\circ}00'12.6''$S    & 2270 & 6.5   &  $2.36'\times2.36'$ \\
\hline
\end{tabular}
\end{center}
%%% for Yolande
%}
%%% for Yolande

\label{tbl:table1}
\end{table}

\clearpage

%%%% table2
%{\large Table2}
%\bigskip

%%% \setlength{\parindent}{-50pt}

\begin{longtable}{lrrrcc}
\caption{Aspect data of observed asteroids.
Date of observations (mid-time of the observing night) in UT,
ecliptic longitude $\lambda$ (deg),
ecliptic latitude $\beta$ (deg),
solar phase angle $\alpha$ (deg), and
abbreviated codes of the observatories
(S: Steward,                          % 1
 V: Vatican,                          % 2
 M: Maidanak,                         % 3
 L: Lulin,                            % 4
 K: Kiso, and                         % 6
 F: Fukuoka),                         % 5
and the sky condition at the observational night
(P: photometric, NP: non-photometric).
% Note that there seems no local extremum in the lightcurves
% obtained from the data denoted as NP${}^\ast$
% (see \ref{subsec:A_3} for more detail).
} \\
%%% for Yolande
%\comment{
%%% for Yolande
\hline
\hline
\multicolumn{1}{c}{Date (UT)} &
 \multicolumn{1}{c}{$\lambda$} &
  \multicolumn{1}{c}{$\beta$}   &
   \multicolumn{1}{c}{$\alpha$}  &
    \multicolumn{1}{c}{obs.}      &
     \multicolumn{1}{c}{cond.} \\
\hline
\hline
\multicolumn{6}{c}{(832) Karin} \\
% 2003--07--31.65 & 334.8     &  1.5    &  9.80 & F & NP \\  % not used
% 2003--08--01.64 & 334.7     &  1.5    &  9.42 & F & NP \\  % not used
% 2003--08--02.69 & 334.5     &  1.5    &  9.02 & F & NP \\  % not used
% 2003--08--03.73 & 334.3     &  1.5    &  8.62 & F & NP \\  % not used
% 2003--08--06.76 & 333.8     &  1.5    &  7.42 & F & NP \\  % not used
% 2003--08--09.07 & 333.3     &  1.5    &  6.22 & F & NP \\  % not used
2003--08--22.64 & 330.7     &  1.6    &  0.85 & F &  P \\
2003--08--23.64 & 330.5     &  1.6    &  0.61 & F &  P \\
2003--09--03.63 & 328.2     &  1.6    &  4.70 & F &  P \\
2003--09--04.63 & 328.0     &  1.6    &  5.13 & F &  P \\
2003--09--05.63 & 327.8     &  1.6    &  5.54 & K & NP \\
2003--09--26.19 & 324.8     &  1.5    & 13.36 & V &  P \\
2003--09--27.19 & 324.7     &  1.5    & 13.68 & V &  P \\
2003--09--28.17 & 324.6     &  1.5    & 13.99 & V &  P \\
2003--09--29.17 & 334.5     &  1.5    & 14.30 & V &  P \\

\hline
\multicolumn{6}{c}{(7719) 1997 GT${}_{36}$} \\
2003--10--14.16 & 315.5     & 0.9     & 18.39 & S &  P \\
2003--10--15.15 & 315.6     & 0.9     & 18.53 & S &  P \\
2003--10--16.14 & 315.6     & 0.9     & 18.66 & S &  P \\
2003--10--17.14 & 315.7     & 0.8     & 18.79 & S &  P \\

\hline
\multicolumn{6}{c}{(10783) 1991 RB${}_{9}$} \\
2004--03--24.50 & 236.6     & 1.7     & 14.94 & S &  P \\
2004--03--26.46 & 236.0     & 1.8     & 14.51 & S & NP \\
2004--03--27.43 & 235.9     & 1.8     & 14.30 & S & NP \\
2004--05--07.66 & 229.9     & 2.3     &  1.12 & L & NP \\
2004--05--09.76 & 229.4     & 2.3     &  0.76 & L &  P \\
2004--05--10.68 & 229.2     & 2.3     &  0.83 & L &  P \\
% 2004--05--11.63 & 229.0     & 2.3     &  1.04 & L & NP \\  % not used
2004--05--11.81 & 229.0     & 2.3     &  1.09 & M & NP \\
2004--05--13.81 & 228.6     & 2.3     &  1.74 & M & NP \\

\hline
\multicolumn{6}{c}{(11728) Einer} \\
2003--05--08.44 & 251.9     & 3.0     &  8.57 & V &  P \\
2003--05--09.44 & 251.7     & 3.0     &  8.20 & V &  P \\
2003--06--29.62 & 242.3     & 2.1     & 11.97 & L &  P \\
2003--06--30.53 & 242.2     & 2.1     & 12.26 & L &  P \\

\hline
\multicolumn{6}{c}{(13765) Nansmith} \\
2003--09--29.46 & 47.2      & 1.3     & 14.20 & V &  P \\
2003--10--15.38 & 45.1      & 1.4     &  8.53 & S &  P \\
2003--10--16.37 & 45.0      & 1.4     &  8.14 & S &  P \\
2003--10--17.37 & 44.8      & 1.4     &  7.73 & S &  P \\
2003--10--23.46 & 38.5      & 0.9     &  5.10 & K & NP \\
2003--10--24.34 & 38.7      & 0.9     &  4.67 & K &  P \\
2003--10--26.42 & 39.2      & 0.9     &  3.80 & K & NP \\
2003--10--27.43 & 39.4      & 0.9     &  3.36 & K & NP \\
% 2003--12--23.22 & 34.8      & 1.0     & 17.66 & S & NP${}^\ast$ \\  % HEC!

\hline
\multicolumn{6}{c}{(16706) Svojsik} \\
2003--05--08.23 & 187.4     & 2.5     & 12.33 & V &  P \\
2003--05--09.30 & 187.3     & 2.5     & 12.63 & V &  P \\
% 2003--05--10.20 & 187.2     & 2.5     & 12.88 & V & NP${}^\ast$ \\  % HEC!
% 2003--05--11.24 & 187.2     & 2.5     & 13.15 & V & NP${}^\ast$ \\  % HEC!

\hline
\multicolumn{6}{c}{(28271) 1999 CK${}_{16}$} \\
2002--11--17.69 &  64.7    & $-1.2$   &  4.48 & L &  P \\
2002--12--01.71 &  64.6    & $-1.3$   &  1.71 & L & NP \\
% 2002--12--03.70 &  64.2    & $-1.3$   &  2.55 & L & NP${}^\ast$ \\  % HEC!
2002--12--04.63 &  64.0    & $-1.3$   &  2.95 & L &  P \\
2002--12--05.58 &  63.8    & $-1.3$   &  3.36 & L &  P \\
2004--03--24.29 & 161.1    & $-1.3$   &  7.81 & S &  P \\
2004--03--26.27 & 161.0    & $-1.2$   &  8.56 & S & NP \\
2004--03--27.22 & 160.6    & $-1.3$   &  8.90 & S & NP \\

\hline
\multicolumn{6}{c}{(40921) 1999 TR${}_{171}$} \\
2003--07--20.71 & 300.5    & $-3.1$   &  1.62 & L & NP \\
2003--07--21.69 & 300.3    & $-3.1$   &  1.37 & L & NP \\

\hline
\multicolumn{6}{c}{(43032) 1999 VR${}_{26}$} \\
2003--08--01.89 & 342. 8   & $-4.3$   & 12.14 & M &  P \\
2003--08--02.90 & 342.7    & $-4.3$   & 11.79 & M &  P \\
2003--08--03.89 & 342.5    & $-4.3$   & 11.44 & M &  P \\
2003--08--04.86 & 342.4    & $-4.3$   & 11.10 & M &  P \\
%2003--08--19.72 & 159.9   & $-4.5$   &  5.35 & L           % not used ?
2003--09--22.25 & 333.3    & $-4.2$   &  9.32 & V &  P \\
2003--09--27.24 & 332.6    & $-4.1$   & 11.12 & V &  P \\
2003--09--28.18 & 332.5    & $-4.1$   & 11.44 & V &  P \\
2003--09--29.17 & 332.4    & $-4.1$   & 11.78 & V &  P \\

\hline
\multicolumn{6}{c}{(69880) 1998 SQ${}_{81}$} \\
2003--09--22.44 & 20.4     & $-1.7$   &  7.72 & V &  P \\
2003--09--26.47 & 19.6     & $-1.8$   &  6.10 & V &  P \\
2003--09--27.41 & 19.5     & $-1.8$   &  5.71 & V &  P \\
2003--09--28.45 & 19.3     & $-1.8$   &  5.28 & V &  P \\
2003--09--29.34 & 19.1     & $-1.9$   &  4.90 & V &  P \\
2003--10--14.33 & 15.9     & $-2.0$   &  1.84 & S &  P \\

\hline
\multicolumn{6}{c}{(71031) 1999 XE${}_{68}$} \\
2003--09--01.87 & 353.2    & $-2.3$   &  5.05 & M &  P \\
2003--09--02.86 & 353.0    & $-2.4$   &  4.66 & M &  P \\
2003--09--03.85 & 352.8    & $-2.4$   &  4.27 & M &  P \\
2003--09--26.34 & 348.1    & $-2.6$   &  5.21 & V &  P \\
2003--09--28.27 & 347.7    & $-2.7$   &  5.96 & V &  P \\

\hline

%%% for Yolande
%}
%%% for Yolande

\label{tbl:table2}
\end{longtable}

\clearpage

%%%% table3
%{\large Table3}
%\bigskip

\begin{small}

\begin{table}[htbp]
\caption[]{Major observational results.
$P$ is the rotation period (hours),
$A(0)$ is the reduced peak-to-trough variation,
$\alpha$ is the solar phase angle during our observation (deg),
QC is the quality code of the period results,
and
the panel designation in Fig. \protect{\ref{fig:lc_all}}.
For (28271) 1999CK${}_{16}$,
$\ast$    denotes the observation result in 2002, and
$\dagger$ denotes the observation result in 2004.
The $\pm$ errors in $P$ are derived from the stepsize of the
CLEAN analysis.

}
%%% for Yolande
%\comment{
%%% for Yolande
\begin{center}
\begin{tabular}{cccccc}
\hline
Asteroid & $P$ & $A(0)$ & $\alpha$ & QC & Fig. \protect{\ref{fig:lc_all}} \\
\hline
  (832) Karin           & $18.348 {+0.037 \atop -0.037}$ & $0.56 \pm 0.02$ &  0.6--14.3 & 2     & a \\
%(4507) 1990 FV         & $ 6.58 \pm 0.04$ & $0.40 \pm 0.03$ &  2.3--13.7 & 3     &     \\
 (7719) 1997 GT${}_{36}$& $29.555 {+1.231 \atop -1.137} $ & $0.31 \pm 0.02$ & 18.4--18.8 & 2     & b \\
(10783) 1991 RB${}_{9}$ & $ 7.334 {+0.005 \atop -0.004} $ & $0.26 \pm 0.02$ &  0.8--14.9 & 3     & c \\
(11728) Einer	        & $13.622 {+0.150 \atop -0.140} $ & $0.14 \pm 0.01$ &  8.6--12.3 & 2     & d \\
(13765) Nansmith        & $10.526 {+0.014 \atop -0.014} $ & $0.06 \pm 0.02$ &  7.7--17.7 & 2     & e \\
(16706) Svojsik		& $ 5.866 {+0.120 \atop -0.120} $ & $0.09 \pm 0.04$ & 12.3--13.2 & 1     & f \\
(28271)${}^{\ast}$    1999 CK${}_{16}$
                        & $ 5.635 {+0.005 \atop -0.010} $ & $0.07 \pm 0.04$ &  1.7--4.5  & 2     & g \\
(28271)${}^{\dagger}$ 1999 CK${}_{16}$
                        & $ 5.645 {+0.043 \atop -0.043} $ & $0.17 \pm 0.02$ &  7.8--8.9  & 2     & h \\
(40921) 1999 TR${}_{171}$& $6.662 {+0.346 \atop -0.158} $ & $0.35 \pm 0.02$ &  1.4--1.6  & 2     & i \\
(43032) 1999 VR${}_{26}$ &$32.890 {+0.078 \atop -0.077} $ & $0.60 \pm 0.06$ &  9.3--12.1 & 2     & j \\
(69880) 1998 SQ${}_{81}$ & $7.675 {+0.010 \atop -0.014} $ & $0.08 \pm 0.01$ &  1.8--7.7  & 2     & k \\
(71031) 1999 XE${}_{68}$ &$20.187 {+0.064 \atop -0.064} $ & $0.39 \pm 0.04$ &  4.3--6.0  & 2     & l \\
\hline
\end{tabular}
\end{center}
%%% for Yolande
%}
%%% for Yolande
\label{tbl:table3}
\end{table}%

\end{small}

\clearpage
%%%%%%%%%%%
%%%% figure captions
%%%%%%%%%%%
%\subsection*{Figure captions}

\begin{figure}[htbp]
\begin{center}
\includegraphics[width=1.0\linewidth]{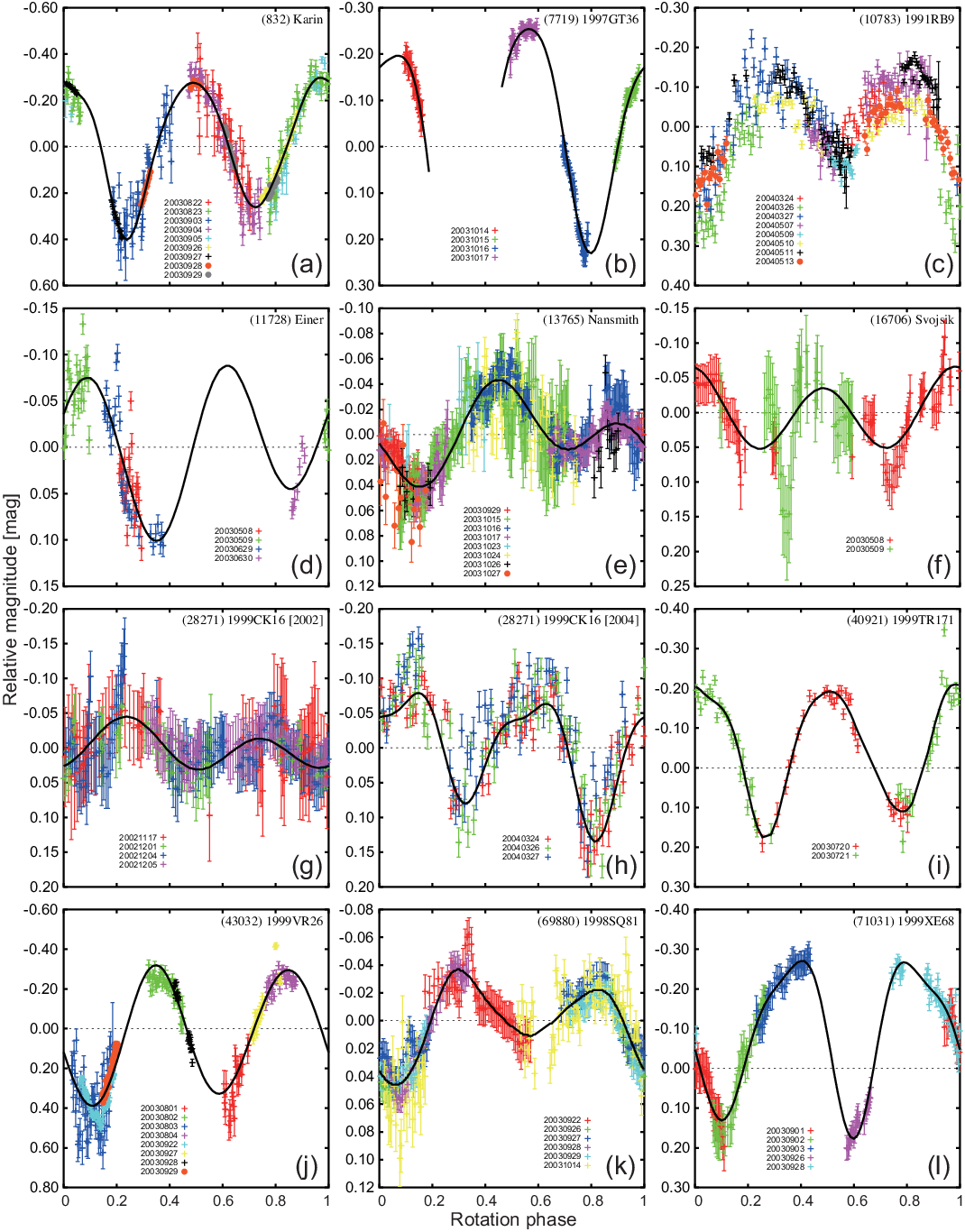}
\caption[]{}
\label{fig:lc_all}
\end{center}
\end{figure}

\begin{figure}[htbp]
\begin{center}
\includegraphics[width=1.0\linewidth]{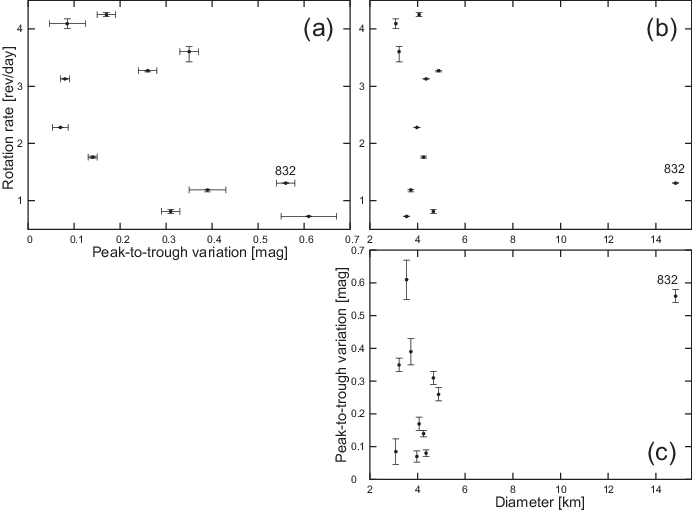}
\caption[]{}
\label{fig:stat}
\end{center}
\end{figure}

\begin{figure}[htbp]
\begin{center}
\includegraphics[width=1.0\linewidth]{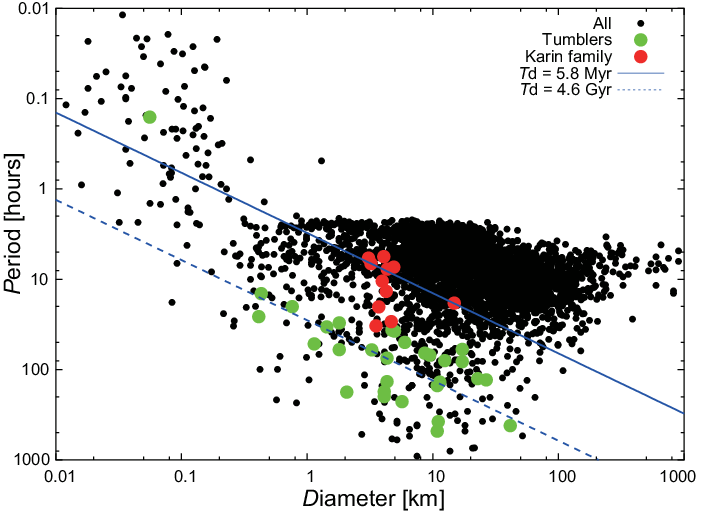}
\caption[]{}
\label{fig:D-P}
\end{center}
\end{figure}

{\large Fig. \ref{fig:lc_all}:}
Results of the lightcurve analysis
of eleven Karin family asteroids.
(a)   (832) Karin,
(b)  (7719) 1997 GT${}_{36}$.
(c) (10783) 1991 RB${}_{9}$,
(d) (11728) Einer
(e) (13765) Nansmith.
(f) (16706) Svojsik,
(g) (28271) 1999 CK${}_{16}$ (observed in 2002),
(h) (28271) 1999 CK${}_{16}$ (observed in 2004),
(i) (40921) 1999 TR${}_{171}$,
(j) (43032) 1999 VR${}_{26}$,
(k) (69880) 1998 SQ${}_{81}$, and
(l) (71031) 1999 XE${}_{68}$.
The vertical axis denotes the relative magnitude referred to a field
star at each observing night.
Note that the lightcurve of (832) Karin in (a) is a
quoted one from \citet{yoshida2004a}.

\bigskip

{\large Fig. \ref{fig:stat}:}
Relation between the rotation rate $1/P$, diameter $D$, and
the reduced peak-to-trough variation $A(0)$ of the
eleven Karin family asteroids that we observed.
(a) $A(0)$ and $1/P$,
(b) $D$ and $1/P$, and
(c) $D$ and $A(0)$.
The largest member, (832) Karin, is denoted as ``832''.
Note that as for (28271) 1999 CK${}_{16}$ we used its values obtained from
our 2004 observation, as its $A(0)$ value in 2004 is larger than
its value in 2002, being closer to their maximum.

\bigskip

{\large Fig. \ref{fig:D-P}:}
Relation between the rotation period $P$ (hours) and diameter $D$ (km) of
3,745 known asteroids (filled black circles)
including 31 tumblers (filled green circles)
and the eleven Karin family asteroids (filled red circles).
The diagonal blue lines show the theoretical $(D,P)$ relation of asteroids when
their damping timescale $T_{\rm d} = 5.8$ Myr (the upper solid blue line)
and when
their damping timescale $T_{\rm d} = 4.6$ Gyr (the lower dashed blue line)
calculated by Eq. (\ref{eqn:damping}).

\newpage
%%%%%%%%%%%
%%%% figures
%%%%%%%%%%%

%\subsection*{Figures}

%%%% new Fig.1

\newpage

%%%% new Fig.2

\newpage

%%%% new Fig.3

% \appendix
% \clearpage

\end{document}